\documentclass{svjour3}
\usepackage{graphicx}
\usepackage{amssymb}
\usepackage{amsmath}
\usepackage{mathptmx}

\smartqed

\begin{document}
\title{Exact solutions of Bianchi I spacetimes which admit Conformal Killing
vectors}

%\titlerunning{Short form of title}        % if too long for running head

\author{Michael Tsamparlis         \and
        Andronikos Paliathanasis   \and
        Leonidas Karpathopoulos.
}

\authorrunning{M. Tsamparlis, A. Paliathanasis, L. Karpathopoulos} % if too long for running head

\institute{M. Tsamparlis \and L. Karpathopoulos \at
              Faculty of Physics, Department of Astrophysics - Astronomy - Mechanics University of Athens, Panepistemiopolis, Athens 157 83, Greece \\
              %\email{mtsampa@phys.uoa.gr}           %  \\
%             \emph{Present address:} of F. Author  %  if needed
           \and
        A. Paliathanasis \at
              Dipartimento di Fisica, Universita' di Napoli, "Federico II" Complesso Universitario di Monte S. Angelo, Via Cintia Edificio 6, I-80126 Napoli, Italy \at
              INFN, Sezione di Napoli, Complesso Universitario di Monte S. Angelo, Via Cintia Edificio 6, I-80126 Napoli, Italy \\
              \email{paliathanasis@na.infn.it}           %  \\
%             \emph{Present address:} of F. Author  %  if needed
       % \and
       %  \at
           %   Faculty of Physics, Department of Astrophysics - Astronomy - Mechanics University of Athens, %Panepistemiopolis, Athens 157 83, Greece \\
           %   \emph{Present address:} of F. Author  %  if needed
}

\date{Received: date / Accepted: date}
% The correct dates will be entered by the editor

\maketitle

\begin{abstract}
We develop a new method in order to classify the Bianchi I spacetimes which
admit conformal Killing vectors (CKV). The method is based on two
propositions which relate the CKVs of 1+(n-1) decomposable Riemannian spaces
with the CKVs of the (n-1) subspace and show that if 1+(n-1) space is
conformally flat then the (n-1) spacetime is maximally symmetric. The method
is used to study the conformal algebra of the Kasner spacetime and other
less known Bianchi type I matter solutions of General Relativity.
% \PACS{PACS code1 \and PACS code2 \and more}
% \subclass{MSC code1 \and MSC code2 \and more}
\end{abstract}

%\titlerunning{Short form of title}        % if too long for running head

\authorrunning{M. Tsamparlis, A. Paliathanasis, L. Karpathopoulos}
% if too long for running head

\institute{M. Tsamparlis \and L. Karpathopoulos \at
              Faculty of Physics, Department of Astrophysics - Astronomy - Mechanics University of Athens, Panepistemiopolis, Athens 157 83, Greece \\
                         \and
        A. Paliathanasis \at
              Dipartimento di Fisica, Universita' di Napoli, "Federico II" Complesso Universitario di Monte S. Angelo, Via Cintia Edificio 6, I-80126 Napoli, Italy \at
              INFN, Sezione di Napoli, Complesso Universitario di Monte S. Angelo, Via Cintia Edificio 6, I-80126 Napoli, Italy \\
              \email{paliathanasis@na.infn.it}                                               }

% The correct dates will be entered by the editor

\section{Introduction}

The Bianchi models are spatially homogeneous spacetimes which admit a group
of motions $G_{3}$ \cite{kramer-stephani,wald} acting on spacelike
hypersurfaces. These spacetimes include the non-isotropic generalizations of
the Friedman-Robertson-Walker (FRW) space-time and have been used in the
discussion of anisotropies in a primordial universe and its evolution
towards the observed isotropy of the present epoch \cite%
{jacobs,narlikar-hoyle-misner}

\noindent The simplest type of these spacetimes are the Bianchi I models for
which $G_{3}$ is the abelian group of translations of the three dimensional
Euclidian space $E^{3}$. In synchronous coordinates the metric of Bianchi I
spacetimes is:
\begin{equation}
ds^{2}=-dt^{2}+A^{2}\left( t\right) dx^{2}+B^{2}\left( t\right)
dy^{2}+C^{2}\left( t\right) dz^{2}  \label{sx1.1}
\end{equation}%
where $A(t),B(t),C(t)$ are functions of the time coordinate only and the
corresponding KVs are $\left\{ \partial _{x},\partial _{y},\partial
_{y}\right\} $. When two of the metric functions are equal, e.g. $%
A^{2}\left( t\right) =B^{2}\left( t\right) $, a Bianchi I spacetime (\ref%
{sx1.1}) reduces to the important class of Locally Rotational Symmetric
(LRS) spacetimes \cite{kramer-stephani}.

A Conformal Killing Vector (CKV) $X^{a}$ is defined by the requirement $%
\mathcal{L}_{X}g_{ab}=2\psi g_{ab}$ and reduces to a Killing vector (KV) ($%
\psi =0$), to a Homothetic Killing Vector (HV) ($\psi _{;a}=0$), and to a
Special Conformal Killing Vector (SCKV) ($\psi _{;ab}=0$). The effects of
these vectors can be seen at all levels of General Relativity, that is,
geometry, kinematics and dynamics. At the geometry level the knowledge of a
CKV makes possible the choice of coordinates so that the metric is
simplified, in the sense that one of the metric components is singled out
\cite{Petrov,Tsamp1}. At the level of kinematics the CKVs impose
restrictions on the kinematic variables (rotation, expansion and shear) and
produce well known results (see for example \cite%
{maartens-mason-tsamparlis,mason-maartens,maartens-maharaj-tupper,coley-tupper1}%
). Finally at the level of dynamics the CKVs can (and have) been used in
various directions, for example to obtain new solutions of the field
equations with (hopefully) better physical properties (see for example \cite%
{coley-tupper1,coley-tupper2,coley-tupper3,coley-tupper4,herrera-leon}). It
becomes evident that it is important that we know the conformal algebra of a
given spacetime.

In \cite{Apost-Tsamp1} all LRS spacetimes which admit CKVs have been
determined. In the following we determine all Bianchi I spacetimes which are
not reducible to LRS\ spacetimes and admit CKVs.

The general Bianchi I spacetime (\ref{sx1.1}) does not admit CKVs. However,
as we will show, there are two families of Bianchi I spacetimes which admit
CKVs. One family consists of the conformally flat Bianchi I spacetimes,
which admit 15 CKVs and are conformally related\footnote{%
These spacetimes are 1+3 spacetimes in which the 3d hypersurface is a
maximally symmetric space with positive and negative curvature scalar
respectively} to Rebou\c{c}as and Tiommo (RT) and Rebou\c{c}as and Teixeira
(ART) \cite{RT,ART}) spacetimes. The second family contains the not
conformally flat Bianchi I spacetimes, which admit only one proper CKV. In
the determination of the CKVs we use the Bilyanov - Defrise - Carter theorem
which relates the conformal algebra of conformally related metrics (for
details see \cite{Defrise-Carter} \cite{Hall-Steele}).

In the literature one finds very few cases of Bianchi I spacetimes which
admit proper CKVs. For example even the CKV found by Maartens and Mellin
\cite{maartens-mellin} is really a CKV in an LRS spacetime and not in a
Bianchi I spacetime \cite{Apost-Tsamp1}. The difficulty lies in the fact
that the direct solution of the conformal equations in Bianchi I spacetimes
is a major task. Thus an alternative simpler method is needed to solve this
problem and this is what it is developed in the following sections. It is to
be noted that using the Petrov classification and the Bilyanov - Defrise -
Carter theorem McIntosh and Steele \cite{McSteele} have determined all
vacuum Bianchi I spacetimes which admit a homothety.

One extra advantage of the proposed method is that one can use it to
prove/test if a given Bianchi I spacetime admits a CKV or not. For example
as it will be shown the two well known anisotropic Bianchi I solutions that
is, the Kasner solution \cite{kramer-stephani} and the anisotropic dust
solution\cite{ellis-hawking}, which have formed the basis of many studies of
anisotropic universes, do not admit a proper CKV; in particular the Kasner
spacetime admits a HV.

The structure of the paper is as follows. In section \ref{preliminaries} we
present two propositions required for the computation of the CKVs in Bianchi
I spacetimes. In sections \ref{classAC} and \ref{classBC} we apply the
results of section \ref{preliminaries} and we determine all Bianchi I
spacetimes which admit CKVs. In section \ref{Exact} we consider the
application of these results in various Bianchi I metrics found in the
literature. Finally in section \ref{Dis} we discuss our results.

\section{Preliminaries}

\label{preliminaries}

As it has been remarked in the last section the computation of CKVs of
Bianchi I spacetimes by direct solution of the conformal equations is a
difficult task. Thus we have developed an indirect method which is based on
the two Propositions discussed below. The first is proposition \ref{prop21}
which has been given in \cite{Tsamp-Nikol-Apost} for spacetimes ($n=4$) and
below is generalized\footnote{%
The generalization of the proof of \cite{Tsamp-Nikol-Apost} to $n$
dimensions is similar to the one for $n=4$ and we omit it.} to $n$-
dimensional Riemannian spaces as follows.

\begin{proposition}
\label{prop21} A decomposable $1+\left( n-1\right)$ ($n\ge 3$ ) Riemannian space %
$g_{ab}$ with line element (Greek indices take the values 1,...,n and Latin indices
the values 0,...,n)%

\begin{equation}
ds^{2}=\varepsilon dt^{2}+h_{\mu \nu }\left( x^{\sigma }\right) dx^{\mu
}dx^{\nu }  \label{sx2.1}
\end{equation}%
admits a proper CKV $X^{a}$ if and only if the $\left( n-1\right) $ space $%
h_{\mu \nu }\left( x^{\sigma }\right) $ admits a gradient proper CKV~$\xi
^{\mu }$. In particular the two vector fields are related as follows%
\begin{equation}
X^{a}=-\frac{\varepsilon }{p}\dot{\lambda}\left( t\right) \psi \left(
x^{\sigma }\right) \partial _{t}+\frac{1}{p}\lambda \left( t\right) \xi
^{\mu }\left( x^{\sigma }\right) +H^{\mu }\left( x^{\sigma }\right)
\label{sx2.0a}
\end{equation}%
where:

- $p$ is a non vanishing constant,

- $\psi \left( x^{\sigma }\right) $ is the conformal factor of the CKV $\xi
^{\mu }$ and satisfies the condition%
\begin{equation}
\psi _{;\mu \nu }=p\psi h_{\mu \nu }  \label{sx2.0b}
\end{equation}%
that is, $\psi _{;\mu}$ is a gradient CKV of the $n-1$
space

- $\lambda \left( t\right) $ satisfies the linear second order equation
\begin{equation}
\ddot{\lambda}\left( t\right) +\varepsilon p\lambda \left( t\right) =0~
\label{sx.02c}
\end{equation}%
where $\dot{\lambda}=\frac{d\lambda }{dt}$.

- $H^{\mu }$ is a KV or a HV of the $n-1$ metric $h_{\mu \nu }\left(
x^{\sigma }\right) .$
\end{proposition}

From proposition \ref{prop21} follows that a proper CKV of the $n-1$ metric $%
h_{\mu \nu }$ generates two proper CKVs for the $n$ metric $g_{ab}$.

\noindent The crucial result of proposition \ref{prop21} is that the
gradient CKVs of the $(n-1)$ space are of the specific form (\ref{sx2.0b}).
Furthermore if the $(n-1)$ space has constant \textit{non vanishing}
Ricciscalar $R$, then the constant $p$ is given by the expression
\begin{equation}
p=\frac{R}{(n-1)(n-2)}\qquad (n\geq 3).  \label{Sx.1.0}
\end{equation}

The second Proposition \ref{prop22} \ concerns the $1+\left( n-1\right) $
decomposable spacetimes which admit CKVs\footnote{%
The proof of Proposition \ref{prop22} is given in appendix \ref{appendixA}}.

\begin{proposition}
\label{prop22} The metric (\ref{sx2.1}) is conformally flat if and
only if the $(n-1)$   metric $h_{\mu \nu }\left( x^{\sigma }\right)
$ is the metric of a space of constant curvature ($n\geq 3$).
\end{proposition}

In addition to these propositions we recall the following result (see \cite%
{Tsamp-Nikol-Apost})

\textit{The metric of a space of constant non-vanishing curvature \textit{of
dimension }$n$\ admits }$n+1$\textit{\ gradient CKVs}.

From proposition \ref{prop22} it follows that as far as the admittance of
CKVs is concerned, the connected $1+(n-1)$ decomposable spaces are
classified in two major classes.

\begin{description}
\item[i)] Class A: The $1+(n-1)$ space is conformally flat. Then the $(n-1)$
space is not conformally flat and the $1+(n-1)~$space admits $\frac{%
(n+1)(n+2)}{2}$ CKVs\newline

\item[ii)] Class B: The $1+(n-1)$ space is not conformally flat. Then the $%
(n-1)$ space is not a space of constant curvature.
\end{description}

\noindent For a space conformally related to a $1+(n-1)$ decomposable space
this classification of CKVs remains the same since all conformally related
spaces admit the same conformal algebra.

\noindent We conclude that the parameter in the classification of the
connected $1+(n-1)$ spacetimes which admit CKVs is the constancy or not of
the curvature scalar of the $(n-1)$ space. Using this observation and
propositions \ref{prop21} and \ref{prop22} we are able to determine all
Bianchi I spacetimes which admit CKVs.

\subsection{CKVs of Bianchi I spacetimes}

\label{CKVsB1}

In the generic line element (\ref{sx1.1}) of Bianchi I spacetime, we
consider the coordinate transformation $dt=C\left( \tau \right) d\tau $ and
get%
\begin{equation}
ds^{2}=C^{2}\left( \tau \right) \left( dz^{2}+ds_{(3)}^{2}\right)
\label{sx4.1}
\end{equation}%
where $ds_{\left( 3\right) }^{2}$ is the three dimensional metric
\begin{equation}
ds_{\left( 3\right) }^{2}=-d\tau ^{2}+B_{1}^{2}(\tau )dy^{2}+A_{1}^{2}(\tau
)dx^{2}  \label{sx4.2}
\end{equation}%
with $A_{1}^{2}\left( \tau \right) =\frac{A^{2}\left( \tau \right) }{%
C^{2}\left( \tau \right) }$,~$B_{1}^{2}\left( \tau \right) =\frac{%
B^{2}\left( \tau \right) }{C^{2}\left( \tau \right) }$. Applying a second
transformation $d\tau =B_{1}^{2}\left( \bar{\tau}\right) d\bar{\tau}$ and $%
\Gamma ^{2}\left( \bar{\tau}\right) =\frac{A_{1}^{2}\left( \bar{\tau}\right)
}{B_{1}^{2}\left( \bar{\tau}\right) }$ the three dimensional metric (\ref%
{sx4.2}) becomes%
\begin{equation}
ds_{\left( 3\right) }^{2}=B_{1}^{2}(\bar{\tau})ds_{1+2}^{2}  \label{sx4.3}
\end{equation}%
where%
\begin{equation}
ds_{1+2}^{2}=dy^{2}+ds_{\left( 2\right) }^{2}  \label{sx4.4}
\end{equation}%
and
\begin{equation}
ds_{\left( 2\right) }^{2}=-d\bar{\tau}^{2}+\Gamma ^{2}(\bar{\tau})dx^{2}.
\label{sx4.5}
\end{equation}%
The two dimensional metric (\ref{sx4.5}) is conformally flat\footnote{%
All $2d$ metrics are conformally flat.}. Indeed if we introduce the new
variable $d\bar{\tau}=\Gamma \left( \hat{\tau}\right) d\hat{\tau}$ the
metric $ds_{\left( 2\right) }^{2}$ becomes%
\begin{equation}
ds_{\left( 2\right) }^{2}=\Gamma ^{2}(\hat{\tau})(-d\hat{\tau}^{2}+dx^{2}).
\label{sx4.6}
\end{equation}

The $2d$ metric $\eta _{AB}=diag\left( -1,1\right) $ admits the three KVs

\begin{equation}
\mathbf{P}_{\hat{\tau}}=\partial _{\hat{\tau}}\qquad \mathbf{P}_{x}=\partial
_{x}\qquad \mathbf{r}=x\partial _{\hat{\tau}}+\hat{\tau}\partial _{x}
\label{sx4.10}
\end{equation}%
and the gradient HV
\begin{equation}
\mathbf{H}=\hat{\tau}\partial _{\hat{\tau}}+x\partial _{x}~,~\psi _{H}=1.
\label{sx4.11}
\end{equation}
The curvature scalar $R_{(2)}$ of the 2d-metric (\ref{sx4.5}) is calculated
to be:
\begin{equation}
R_{\left( 2\right) }=2\frac{\Gamma _{,\bar{\tau}\bar{\tau}}}{\Gamma }.
\label{sx4.14}
\end{equation}

According to proposition \ref{prop22} the condition that the $(1+2)$d -
metric (\ref{sx4.4}) - and consequently the 3d-metric $ds_{3}^{2}$ - is
conformally flat, is that the 2d-metric (\ref{sx4.6}) is a the metric of a
space of constant curvature. We set $R_{(2) }=const.=2c$ and find that that
this is the case when $\Gamma _{,\bar{\tau}\bar{\tau}}=c\Gamma $.

On the other hand when $ds_{3}^{2}$ is of constant curvature then by means
of the inverse of proposition \ref{prop22} the metric $ds_{1+3}^{2}$ is
conformally flat hence the metric $ds^{2}$ is also conformally flat.

We conclude that the classification of Bianchi I spacetimes which admit CKVs
is done in two classes:

\emph{Class A }: Contains all Bianchi I spacetimes which are conformally
flat. According to proposition \ref{prop22} in this case the 3d-metric $%
ds_{\left( 3\right) }^{2}$ is of constant curvature and the form of the
metric functions $A_{1}(\tau )$, $B_{1}(\tau )$ is fixed.

\emph{Class B }: Contains all Bianchi I spacetimes which are not conformally
flat therefore the decomposable metric is not conformally flat. According to
the inverse of proposition \ref{prop22} in this class the 3-d metric $%
ds_{\left( 3\right) }^{2}$ is not the metric of a space of constant
curvature.

In Class B there are two cases to be considered.\newline
Case B1: The 3d-metric $ds_{\left( 3\right) }^{2}$ is not conformally flat
in which case the scalar curvature of the 2d-metric $R_{(2)}\neq const.$%
\newline
Case B2: The 3d-metric $ds_{\left( 3\right) }^{2}$ is conformally flat hence
according to proposition \ref{prop22} the 2d-metric $ds_{\left( 2\right)
}^{2}$ is of constant curvature i.e. $R_{(2)}=const.$\newline

In the following we consider each Class and derive the corresponding Bianchi
I spacetimes together with the CKV(s). We ignore the cases $%
A_{1}=B_{1}\Leftrightarrow \Gamma =const.$ which lead to LRS spacetimes
whose CKVs have already been found in \cite{Apost-Tsamp1}.

\section{Class A: The conformally flat Bianchi I spacetimes}

\label{classAC}

Demanding that the Weyl tensor of the metric (\ref{sx4.1}) vanishes we find
the following conditions on the metric functions $A_{1},B_{1}$:

\begin{eqnarray}
A_{1}\ddot{B}_{1}+B_{1}\ddot{A}_{1}-2\dot{A}_{1}\dot{B}_{1} &=&0
\label{sx4.14a1} \\
A_{1}\ddot{B}_{1}-2B_{1}\ddot{A}_{1}+\dot{A}_{1}\dot{B}_{1} &=&0
\label{sx4.14a2} \\
\ddot{A}_{1}B_{1}-2A_{1}\ddot{B}_{1}+\dot{A}_{1}\dot{B}_{1} &=&0
\label{sx4.14a3}
\end{eqnarray}
where a dot over a symbol denotes differentiation with respect to coordinate
$\tau $. We note that only two of these three equations are independent.

Using~(\ref{sx4.14a1})-(\ref{sx4.14a3}) we can prove that the 3-metric (\ref%
{sx4.2}) is the metric of a 3-space of constant curvature $%
R_{(3)}=6\varepsilon a^{2}$ where $\varepsilon =\pm 1$ and $a\ne 0$ is a
constant. There are only two such spacetimes the RT spacetime \cite{RT} and
the ART spacetime \cite{ART} mentioned above.

The RT and the ART spacetimes in isochronous coordinates have the line
element%
\begin{equation}
ds_{RT}^{2}=-dt^{2}+\sin ^{2}(t/a)dx^{2}+\cos ^{2}(t/a)dy^{2}+dz^{2}
\label{sx3.3}
\end{equation}%
\begin{equation}
ds_{ART}^{2}=-dt^{2}+\sinh ^{2}(t/a)dx^{2}+\cosh ^{2}(t/a)dy^{2}+dz^{2}
\label{sx3.4}
\end{equation}%
respectively. These spacetimes are 1+3 decomposable spaces whose three
dimensional space is a space of constant curvature. They admit a 15
dimensional conformal algebra with a seven dimensional Killing subalgebra,
which has been given in \cite{Apost-Tsamp1}. For the completeness of the
paper in appendix \ref{appendixB} we give the conformal algebra of the RT
and the ART spacetimes in a convenient form.

\section{Class B: The non-conformally flat Bianchi I spacetimes}

\label{classBC}

In this class there are two subcases to be considered depending on $%
R_{\left( 2\right) }=const$ and $R_{\left( 2\right) }\neq const$ where $%
R_{\left( 2\right) }$ is the Ricciscalar of the two dimensional space (\ref%
{sx4.6}).

\subsection{Case B.I: $R_{\left( 2\right) }\neq const.$}

In this case we are interested only for the KVs and the HV of $ds_{\left(
2\right) }^{2}$ since if there exist a proper CKV which satisfy condition (%
\ref{sx2.0b}) of Proposition \ref{prop21}, then the two dimensiona space is
of constant curvature. From the CKVs of $d\hat{s}_{\left( 2\right) }^{2}=(-d%
\hat{\tau}^{2}+dx^{2})$ only the ones which do not contain terms $f(\hat{\tau%
})g(x)\partial _{\hat{\tau}}$ with $f(\hat{\tau})\neq \hat{\tau}$ can
satisfy this property. \ It is well known that the two dimensional space $d%
\hat{s}_{\left( 2\right) }^{2}$ admits infinity CKVs. However, the vector
fields which do not contain the terms $f(\hat{\tau})g(x)\partial _{\hat{\tau}%
}$ with $f(\hat{\tau})\neq \hat{\tau}$ are the two vector fields $\mathbf{P}%
_{\hat{\tau}}$ and the $\mathbf{H}$.

The conformal factor of $\mathbf{P}_{\hat{\tau}}$ of the metric $ds_{\left(
2\right) }^{2}$ is:
\begin{equation*}
\psi (\mathbf{P}_{\hat{\tau}})=\Gamma _{,\bar{\tau}}.
\end{equation*}

If we demand $\psi (\mathbf{P}_{\hat{\tau}})=0$ (the case of KVs) then we
get $A_{1}^{2}=B_{1}^{2}$, i.e.$~$the LRS case which we ignore. If we demand
$\psi (\mathbf{P}_{\hat{\tau}})=const$ then we find $\Gamma _{,\bar{\tau}%
\bar{\tau}}=0$ which implies by (\ref{sx4.14}) that $R_{\left( 2\right) }=0$
i.e~constant which contradicts our assumption. Therefore $\mathbf{P}_{\hat{%
\tau}}$ produces nothing relevant.

The HV $\mathbf{H~}$ has conformal factor
\begin{equation}
\psi (\mathbf{H})=\Gamma _{,\bar{\tau}}\int \frac{d\bar{\tau}}{\Gamma }+1.
\label{sx4.15}
\end{equation}%
The requirement that $\mathbf{H}$ is a KV of the 2-metric $ds_{\left(
2\right) }^{2}$ gives $\hat{\tau}\Gamma _{,\hat{\tau}}+\Gamma =0,$ hence $%
\Gamma \left( \hat{\tau}\right) =\frac{\Gamma _{0}}{\hat{\tau}}$ which
implies $R_{\left( 2\right) }=const.$ and it is excluded. The requirement
that $\mathbf{H}$ is a HV with conformal factor $\alpha _{2}(\neq 0)$ gives:
\begin{equation}
\Gamma =c_{1}\hat{\tau}^{\alpha _{2}-1}  \label{sx4.16}
\end{equation}%
where $c_{1}=const$. This HV is acceptable provided that $\alpha _{2}\neq 1$
in order to avoid the LRS case. By proposition \ref{prop21} this gives the
following HV for the 1+2 metric (\ref{sx4.4}):
\begin{equation}
\mathbf{H}_{1}=\alpha _{2}y\partial _{y}+\hat{\tau}\partial _{\hat{\tau}%
}+x\partial _{x}  \label{sx4.17}
\end{equation}%
with conformal factor
\begin{equation}
\psi (\mathbf{H}_{1})=\alpha _{2}.  \label{sx4.18}
\end{equation}

This vector is a non-gradient CKV for the metric (\ref{sx4.3}) with
conformal factor:
\begin{equation}
\bar{\psi}(\mathbf{H}_{1})=\hat{\tau}(\ln A_{1})_{,\hat{\tau}}+\alpha _{2}.
\label{sx4.19}
\end{equation}%
We are interested in KVs and HVs (we show in the Appendix that the gradient
CKVs of the form $\lambda (\mathbf{\xi })_{|\alpha \beta }=p\lambda (\mathbf{%
\xi })g_{\alpha \beta }$ imply that the 3-metric (\ref{sx4.3}) is of
constant curvature) thus we examine possible reductions of this CKV to a KV
or a HV.

If $\mathbf{H}_{1}$ is a KV then $\bar{\psi}(\mathbf{H}_{1})=0$ and this
gives $A_{1}=c_{2}\hat{\tau}^{-\alpha _{2}}$. From (\ref{sx4.6}) and (\ref%
{sx4.16}) we obtain $B_{1}=\frac{c_{1}c_{2}}{\hat{\tau}}$ which implies $%
\hat{\tau}=c_{3}e^{\tau /c}$ where $c=c_{1}c_{2}$. Thus we have the
following KV :
\begin{equation}
X_{B_{1}}=\alpha _{2}y\partial _{y}+c\partial _{\tau }+x\partial _{x}
\label{sx4.20}
\end{equation}%
for the three dimensional metric:
\begin{equation}
ds_{\left( 3\right) }^{2}=-d\tau ^{2}+c_{2}^{2}c_{3}^{-2\alpha
_{2}}e^{-2\alpha _{2}\tau /c}dy^{2}+\left( \frac{c}{c_{3}}\right)
^{2}e^{-2\tau /c}dx^{2}.  \label{sx4.21}
\end{equation}%
Due to proposition \ref{prop21} this is also a KV for the metric $%
ds_{1+3}^{2}=dz^{2}+ds_{(3) }^{2}$ hence a proper CKV for the metric (\ref%
{sx4.1}) with conformal factor (note that $\partial _{\tau }=A_{1}\partial
_{t}$)
\begin{equation}
\psi (X_{B_{1}})=c(C\left( t\right) )_{,t}.  \label{sx4.22}
\end{equation}%
The metric $ds^{2}$ is given in (\ref{sx4.1}) and describes a family of
Bianchi I metrics parameterized by the function $C(t)$.

When $\mathbf{H}_{1}$ is a HV from equation (\ref{sx4.19}) we obtain ($%
\alpha _{3}=const.$):
\begin{equation}
\hat{\tau}(\ln A_{1})_{,\hat{\tau}}+\alpha _{2}=\alpha _{3}\Leftrightarrow
A_{1}=c_{2}\hat{\tau}^{\alpha _{3}-\alpha _{2}}  \label{sx4.23}
\end{equation}%
and
\begin{equation}
B_{1}=c_{1}c_{2}\hat{\tau}^{\alpha _{3}-1}  \label{sx4.24}
\end{equation}%
therefore we have that
\begin{equation}
\hat{\tau}=\left( \frac{\alpha _3}{c_1c_2}\right) ^{1/\alpha _3}\tau
^{1/\alpha _3}.  \label{sx4.25}
\end{equation}
Eventually we have the CKV:
\begin{equation}
X_{B_{1}}=\alpha _{2}y\partial _{y}+\alpha _{3}\tau \partial _{\tau
}+x\partial _{x}+\alpha _{3}z\partial _{z}  \label{sx4.26}
\end{equation}%
for the Bianchi I metric:
\begin{equation}
ds^{2}=C^{2}(\tau )\left[ dz^{2}-d\tau ^{2}+c_{2}^{2}\left( \frac{\alpha _{3}%
}{c_{1}c_{2}}\right) ^{2\frac{(\alpha _{3}-\alpha _{2})}{\alpha _{3}}}\tau
^{2\frac{(\alpha _{3}-\alpha _{2})}{\alpha _{3}}}dy^{2}+c_{1}^{2}c_{2}^{2}%
\left( \frac{\alpha _{3}}{c_{1}c_{2}}\right) ^{2\frac{(\alpha _{3}-1)}{%
\alpha _{3}}}\tau ^{2\frac{(\alpha _{3}-1)}{\alpha _{3}}}dx^{2}\right]
\label{sx4.27}
\end{equation}
with conformal factor:
\begin{equation}
\psi (X_{B_{2}})=\alpha _{3}\left[ 1+\tau (\ln \left\vert C\right\vert
)_{,\tau }\right] .  \label{sx4.28}
\end{equation}

\subsection{Case B.II: $R_{\left( 2\right) }=const$}

We consider the subcases: $R_{\left( 2\right) }=0$, and $R_{\left( 2\right)
}\neq 0$.

When $R_{\left( 2\right) }=0$ from (\ref{sx4.14}) we have that
\begin{equation}
\Gamma =b_{0}\bar{\tau}\Leftrightarrow B_{1}=b_{0}\bar{\tau}A_{1}.
\label{sx4.29}
\end{equation}%
Equation (\ref{sx4.29}) implies that the 3-metric (\ref{sx4.4}) has the form
(we ignore the unimportant integration constant $b_{0}$):
\begin{equation}
ds_{1+2}^{2}=dy^{2}-d\bar{\tau}^{2}+\bar{\tau}^{2}dx^{2}.  \label{sx4.30}
\end{equation}%
The CKVs of the flat 3-metric $ds^{2}=-d\tilde{t}^{2}+d\tilde{x}^{2}+d\tilde{%
y}^{2}$ are known \cite{choquet-bruhat}. Using the transformation $\tilde{t}=%
\bar{\tau}\cosh x,\tilde{x}=\bar{\tau}\sinh x,\tilde{y}=y$ we obtain the
3-metric (\ref{sx4.30}) from which we obtain the following conformal algebra
(we ignore the KVs $\partial _{x},\partial _{y};$ $i=1,2,3,4;$ $\alpha
=1,2,3 $.):

- Four KVs%
\begin{equation*}
\mathbf{X}_{1}=\cosh x\partial _{\bar{\tau}}-\frac{\sinh x}{\bar{\tau}}%
\partial _{x}~~
\end{equation*}%
\begin{equation*}
\mathbf{X}_{2}=\sinh x\partial _{\bar{\tau}}-\frac{\cosh x}{\bar{\tau}}%
\partial _{x}
\end{equation*}%
\begin{equation*}
\mathbf{X}_{3}=y\sinh x\partial _{\bar{\tau}}-y\frac{\cosh x}{\bar{\tau}}%
\partial _{x}+\bar{\tau}\sinh x\partial _{y}
\end{equation*}%
\begin{equation*}
\mathbf{X}_{4}=y\cosh x\partial _{\bar{\tau}}-y\frac{\sinh x}{\bar{\tau}}%
\partial _{x}+\bar{\tau}\cosh x\partial _{y}
\end{equation*}

- one gradient HV%
\begin{equation}
\mathbf{X}_{7}=\bar{\tau}\partial _{\bar{\tau}}+y\partial _{y}~,~\psi (%
\mathbf{X}_{7})=1
\end{equation}%
- three special CKVs%
\begin{equation*}
\mathbf{X}_{8}=(y^{2}+\bar{\tau}^{2})\cosh x\partial _{\bar{\tau}}+\frac{%
\bar{\tau}^{2}-y^{2}}{\bar{\tau}}\sinh x\partial _{x}+2y\bar{\tau}\cosh
x\partial _{y}
\end{equation*}%
\begin{equation*}
\mathbf{X}_{9}=(y^{2}+\bar{\tau}^{2})\sinh x\partial _{\bar{\tau}}+\frac{%
\bar{\tau}^{2}-y^{2}}{\bar{\tau}}\cosh x\partial _{x}+2y\bar{\tau}\sinh
x\partial _{y}
\end{equation*}%
\begin{equation*}
\mathbf{X}_{10}=2\bar{\tau}y\partial _{\bar{\tau}}+(y^{2}+\bar{\tau}%
^{2})\partial _{y}
\end{equation*}

with corresponding conformal factors:
\begin{equation}
\psi \left( \mathbf{X}_{8}\right) =2\bar{\tau}\cosh x~,~\psi \left( \mathbf{X%
}_{9}\right) =2\bar{\tau}\sinh x~,~\psi \left( \mathbf{X}_{10}\right) =2y
\label{sx4.34}
\end{equation}

These vectors are also CKVs for the metric (\ref{sx4.3}) but with conformal
factors:
\begin{equation}
\psi ^{\prime }(\mathbf{X}_{A})=\mathbf{X}_{A}(\ln A_{1})+\psi (\mathbf{X}%
_{A})  \label{sx4.35}
\end{equation}%
where $A=1,2,...,10$. The possible vectors $\mathbf{X}_{A}$ which give $\psi
^{\prime }(\mathbf{X}_{A})=const.$ are the KVs and the HV which do not
contain terms of $f(\bar{\tau})g(x)\partial _{\bar{\tau}}$. The only such
vector is the HV $\mathbf{X}_{7}$.

The case that $\mathbf{X}_{7}$ is a KV for the metric (\ref{sx4.3}) gives $%
B_{1}=const.$ and we ignore it. We set $\psi ^{\prime }(\mathbf{X}%
_{A})=\alpha _{4}$ and we obtain, after standard calculations, that the
vector $\mathbf{X}_{7}=\alpha _{4}\tau \partial _{\tau }+y\partial _{y}$ is
a HV for the 3-metric:
\begin{equation}
ds^{2}=-d\tau ^{2}+b_{1}^{2}\left( \frac{\alpha _{4}}{b_{1}}\right) ^{2\frac{%
\alpha _{4}-1}{\alpha _{4}}}\tau ^{2\frac{\alpha _{4}-1}{\alpha _{4}}%
}dy^{2}+\alpha _{4}^{2}\tau ^{2}dx^{2}  \label{sx4.36}
\end{equation}%
with conformal factor $\alpha _{4}$. This vector is extended to a HV for the
1+3 metric $ds_{1+3}^{2}=dz^{2}+ds_{(3)}^{2}$ which is of the form:
\begin{equation}
X_{B_{3}}=\alpha _{4}\tau \partial _{\tau }+y\partial _{y}+\alpha
_{4}z\partial _{z}.  \label{sx4.37}
\end{equation}%
The Bianchi I metric (\ref{sx4.36}) and the CKV (\ref{sx4.37}) are obtained
from the metric (\ref{sx4.27}) and the CKV (\ref{sx4.26}) if we set $a_{1}=0$
and interchange the coordinates $x,y.$ Therefore it is not a new case. \

A detailed study of the subcase $R_{\left( 2\right) }\neq 0$ shows that
there are no more new Bianchi I metrics which admit CKVs. The calculations
are rather standard and similar to the ones above and are omitted.

We conclude that there are two families of metrics in B.II class
parameterized by the function $C(\tau )$. Each family admits one proper CKV
and have as follows:

Metrics B$_{1}$~with $(\alpha _{1}\neq 0,1~,~c\neq 0)~$

\begin{equation}
ds^{2}=C^{2}(\tau )\left[ -d\tau ^{2}+e^{-\frac{2}{c}\tau }dx^{2}+e^{-\frac{%
2\alpha _{1}}{c}\tau }dy^{2}+dz^{2}\right]  \label{sx3.21}
\end{equation}%
and corresponding CKV

\begin{equation}
X_{B_{1}}=c\partial _{\tau }+x\partial _{x}+\alpha _{1}y\partial _{y}
\label{sx3.22}
\end{equation}

\begin{equation}
\psi (X_{B_{1}})=c\left( \ln \left\vert C\right\vert \right) _{,\tau }
\label{sx3.23}
\end{equation}

Metrics B$_{2}$ with $(\alpha _{2}\neq 0,1)$ and $(\alpha _{1}\neq \alpha
_{2})$

\begin{equation}
ds^{2}=C^{2}(\tau )\left[ -d\tau ^{2}+\tau ^{2\frac{\alpha _{2}-1}{\alpha
_{2}}}dx^{2}+\tau ^{2\frac{\alpha _{2}-\alpha _{1}}{\alpha _{2}}%
}dy^{2}+dz^{2}\right]  \label{sx3.24}
\end{equation}%
and corresponding CKV
\begin{equation}
X_{B_{2}}=\alpha _{2}\tau \partial _{\tau }+\alpha _{1}y\partial
_{y}+x\partial _{x}+\alpha _{2}z\partial _{z}  \label{sx3.25}
\end{equation}
with conformal factor
\begin{equation}
\psi (X_{B_{2}})=\alpha _{2}\left[ 1+\tau (\ln \left\vert C\right\vert
)_{,\tau }\right].  \label{sx3.26}
\end{equation}

We observe that the CKV $X_{B_{1}}$ of the metric B$_{1}$ becomes a HV when $%
\left( \ln \left\vert C\right\vert \right) _{,\tau }=\psi _{0}$, i.e. $%
C\left( \tau \right) =e^{\psi _{0}\tau }$. In that case the metric (\ref%
{sx3.21}) becomes~($e^{\psi _{0}\tau }=t$)
\begin{equation}
ds^{2}=-dt^{2}+t^{-\frac{2}{c\psi _{0}}}dx^{2}+t^{-\frac{2a_{1}}{c\psi _{0}}%
}dy^{2}+t^{2}dz  \label{sx3.27}
\end{equation}%
where we substitute $e^{\psi _{0}\tau }=t.$ Furthermore the metric $B_{2}$
admits a HV when $C\left( \tau \right) =\tau ^{\psi _{0}-1}$. In that case
the line element (\ref{sx3.24}) becomes%
\begin{equation}
ds^{2}=-dt^{2}+t^{2\frac{\alpha _{2}-1}{\psi _{0}\alpha _{2}}}dx^{2}+t^{2%
\frac{\alpha _{2}-\alpha _{1}}{\psi _{0}\alpha _{2}}}dy^{2}+t^{2\frac{\left(
\psi _{0}-1\right) }{\psi _{0}}}dz^{2}  \label{sx3.28}
\end{equation}

Therefore from the spacetimes (\ref{sx3.27}) and (\ref{sx3.28}) we have that
the Bianchi I spacetimes (\ref{sx1.1}) which admit a proper HV are the
spacetimes with power law coefficients. As it has been noted in the
introduction all vacuum Bianchi I spacetimes which admit a Homothetic vector
have been determined in \cite{McSteele}.

In the following section we study the CKVs of some well known exact
solutions of Einstein field equations in a Bianchi I spacetime.

\section{Exact Bianchi I solutions and conformal symmetries}

\label{Exact}

One can apply the results of the last section to determine if a given
Bianchi I metric admits or not CKVs and at the same time determine the exact
form of the CKVs and their conformal factors. The method of work is simple
and consists of the following steps.

From the given Bianchi metric one computes the traceless projection tensor $%
\Delta _{ab}^{cd}=g_{ab}g^{cd}-\frac{1}{4}\delta _{a}^{c}\delta _{b}^{d}$
and demands that $\Delta _{ab}^{cd}X_{c;d}=0$ where $X_{c}$ is any of the
CKVs defined in (\ref{sx3.6}), (\ref{sx3.8}), (\ref{sx3.13}), (\ref{sx3.15})
(conformally flat case) and (\ref{sx3.22}), (\ref{sx3.25}) (non-conformally
flat case). If this condition cannot be satisfied for any values of the
parameters of the metric then the metric does not admit a CKV otherwise it
does. It is possible that the conformal factors are constants in which case
the CKVs reduce to HVs.

Before one proceeds with the above it is convenient to compute the Weyl
tensor and examine if the space is conformally flat or not. If it is not
there is no need to consider the vectors (\ref{sx3.6}), (\ref{sx3.8}), (\ref%
{sx3.13}), (\ref{sx3.15}) whereas if it is there is no need to consider the
vectors (\ref{sx3.22}), (\ref{sx3.25}).

In the following section we apply the above method to various anisotropic
Bianchi I metrics which we have traced in the literature. We present the
derivation of the results for the Kasner type metrics in some detail whereas
the for rest of the metrics we give only the results of the calculations.

\subsection{Kasner type metrics}

The Kasner type metrics are defined by the line element:
\begin{equation}
ds^2=-dt^2+t^{2p}dx^2+t^{2q}dy^2+t^{2r}dz^2  \label{sx3.30}
\end{equation}
where $p,q,r\,$ are different constants (otherwise the metric reduces to an
LRS metric (two of the constants equal) or to a FRW metric (all constants
equal). The well known Kasner spacetime - which has been used extensively in
the literature in the discussion of anisotropies of the Universe - is a
vacuum solution of Einstein's field equations with the parameters $p,q,r$
restricted by the relations:
\begin{eqnarray}
p+q+r &=&1  \label{sx3.31} \\
p^2+q^2+r^2 &=&1.  \notag
\end{eqnarray}

\noindent Kasner spacetime is vacuum so if conformally flat it is flat
therefore we have a non-conformally flat case. Condition $\Delta
_{ab}^{cd}X_{c;d}=0$ for the vector fields (\ref{sx3.22}),(\ref{sx3.25})
yields in turn:

\underline{$X_{B_1}$}:

\noindent We find $r=1,$ $c=1,\alpha _{1}=\frac{q-1}{p-1},$ $\tau d\tau =$ $%
\frac{1}{p-1}tdt$ ($p\neq 1$ otherwise we have an LRS spacetime) from which
follows that the Kasner type metric:
\begin{equation}
ds^2=-dt^2+t^{2p}dx^2+t^{2q}dy^2+t^2dz^2  \label{sx3.32}
\end{equation}
admits the HV \cite{kramer-stephani}:
\begin{equation}
X_{B_{1}}=\frac{1}{1-p}t\partial _{t}+\frac{q-1}{p-1}y\partial
_{y}+x\partial _{x};\;\psi (X_{B_{1}})=\frac{1}{1-p}  \label{sx3.33}
\end{equation}

\underline{$X_{B_2}$}:

We find $r\neq 1,\alpha _{1}=\frac{q-1}{p-1},$ $\alpha _{2}=\frac{r-1}{p-1}$
($p\neq 1)$ from which we conclude that the Kasner type metric (\ref{sx3.30}%
) with $r\neq 1,p\neq 1$ admits the HV:
\begin{equation}
X_{B_{2}}=\frac{r-1}{p-1}\tau \partial _{\tau }+x\partial _{x}++\frac{q-1}{%
p-1}y\partial _{y}+\frac{q-1}{p-1}z\partial _{z};\;\psi (X_{B_{2}})=\frac{1}{%
1-p}\text{ }  \label{sx3.34}
\end{equation}%
We emphasize that due to conditions (\ref{sx3.31}) the Kasner spacetime
admits only the HV $X_{B_{2}}$. These results agree with those of \cite%
{McSteele}.

\subsection{Bianchi I shear free spacetimes}

This class contains many well known solutions of the field equations. The
general form of the spacetime metric is
\begin{equation}
ds^{2}=-dt^{2}+S^{2}(t)f^{2p}(t)dx^{2}+S^{2}(t)f^{2q}(t)dy^{2}+S^{2}(t)f^{2r}(t)dz^{2}
\label{sx3.35}
\end{equation}%
where the functions $S(t),f(t)$ are general functions. The various known
solutions of this form are perfect fluid solutions with vanishing and
non-vanishing cosmological constant $\Lambda $. These solutions are:

a. Dust solution

$\Lambda =0$ \cite{ellis-hawking}.

\begin{equation}
S^3(t)=\frac 92Mt(t+\Sigma );f(t)=\frac{t^{2/3}}{S(t)};p=2\sin \alpha
,q=2\sin (\alpha +\frac{2\pi }3),r=2\sin (\alpha +\frac{4\pi }3)
\label{sx3.36}
\end{equation}
The constant $\alpha $ is the angle where the anisotropy is maximal ($-\frac
\pi 2<\alpha <\frac \pi 2$) and $\Sigma ,M$ are constants with $\Sigma >0$.

$\Lambda \neq 0$ \cite{kramer-stephani}

\begin{equation}
S^{3}(t)=\left\{
\begin{array}{c}
a\sinh \omega t+\frac{M}{2\Lambda }(\cosh \omega t-1)\qquad \text{for}\qquad
\Lambda >0 \\
a\sin \omega t+\frac{M}{2\Lambda }(\cos \omega t-1)\qquad \text{for}\qquad
\Lambda <0%
\end{array}%
\right\}  \label{sx3.37}
\end{equation}

\begin{equation}
f(t)=\left\{
\begin{array}{c}
\frac{\cosh \omega t-1}{S^{3}(t)}\qquad \text{for}\qquad \Lambda >0 \\
\frac{1-\cos \omega t}{S^{3}(t)}\qquad \text{for}\qquad \Lambda <0%
\end{array}%
\right\} .  \label{sx3.38}
\end{equation}

b. Perfect fluid solutions with an equation of state $p=(\gamma -1)\mu ~$%
\cite{jacobs},\cite{kramer-stephani}

\begin{equation}
S^{3}(t)=\left\{
\begin{array}{c}
c\sinh \omega t\qquad \text{for}\qquad \Lambda >0 \\
\sqrt{3(3+M)}t\qquad \text{for}\qquad \Lambda =0 \\
c\sin \omega t\qquad \text{for}\qquad \Lambda <0%
\end{array}%
\right\}  \label{sx3.39}
\end{equation}

\begin{equation}
f(t)=\left\{
\begin{array}{c}
\left( \tanh \frac{\omega t}{t}\right) ^{b}\qquad \text{for}\qquad \Lambda >0
\\
t^{b}\qquad \text{for}\qquad \Lambda =0 \\
\left( \tan \frac{\omega t}{t}\right) ^{b}\qquad \text{for}\qquad \Lambda <0%
\end{array}%
\right\}  \label{sx3.40}
\end{equation}%
where $b=\left( \frac{3}{3+M}\right) ^{1/2}$ and $c=\left( \frac{3+M}{%
\Lambda }\right) ^{1/2}$.

For $\mu =0$ we take the vacuum solutions for $\Lambda =,>,<0$. In \cite%
{kramer-stephani} one can find the form of the solutions for various values
of $\gamma .$

\subsection{Einstein-Maxwell solutions}

We have found two solutions describing cosmological models with an
electromagnetic field satisfying the Rainich conditions. These are:

Data solution \cite{Datta}:

\begin{equation}
ds^2=A^{-1}(-dt^2+A^2dx^2+ABdy^2+ACdz^2)  \label{sx3.41}
\end{equation}
where:

\begin{eqnarray*}
A &=&c_{1}t^{\mu }+c_{2}t^{-\mu } \\
AB &=&t^{\lambda }\quad \text{and}\quad AC=t^{2-\lambda }
\end{eqnarray*}%
and $c_{1},c_{2},\mu ,\lambda $ are constants with $c_{1}c_{2}\neq 0$.

Rosen solution~\cite{Rosen}:

\begin{equation}
ds^2=-\frac{b_1^2(\tan \frac 12t)^{2(b_2+b_3)}}{\sin ^4t}dt^2+\sin ^2tdx^2+%
\frac{(\tan \frac 12t)^{2b_2}}{\sin ^2t}dy^2+\frac{(\tan \frac 12t)^{2b_3}}{%
\sin ^2t}dz^2  \label{sx3.42}
\end{equation}
where $b_1,b_2,b_3$ are constants and $b_2b_3=1$.

\noindent Using the criterion $\Delta _{ab}^{cd}X_{c;d}=0$ for each of the
above spacetimes, we find, after standard but lengthy computations, the
results of Table \ref{Table1}.

%TCIMACRO{\TeXButton{B}{\begin{table}[tbp] \centering}}%
%BeginExpansion
\begin{table}[tbp] \centering%
%EndExpansion
\caption{Exact solutions of Bianchi I spacetimes which admit CKVs}%
\begin{tabular}{ccc}
\hline\hline
\textbf{Spacetime} & \textbf{CKVs} & \textbf{Conformal factor} \\ \hline
Datta Solution & $\nexists $ & $\nexists $ \\
Rosen Solution & $\nexists $ & $\nexists $ \\
Kasner-type & (\ref{sx3.33})~/~(\ref{sx3.34}) & constant \\
Shear free spacetimes &
\begin{tabular}{l}
$\nexists $\quad for $\Lambda >0$ \\
~(\ref{sx3.33})~/~(\ref{sx3.34}) \\
$\nexists $\quad for $\Lambda <0$%
\end{tabular}
&
\begin{tabular}{l}
$\nexists $\quad for $\Lambda >0$ \\
~~~constant \\
$\nexists $\quad for $\Lambda <0$%
\end{tabular}
\\
Dust solution & $\nexists $ & $\nexists $ \\ \hline\hline
\end{tabular}%
\label{Table1}%
%TCIMACRO{\TeXButton{E}{\end{table}}}%
%BeginExpansion
\end{table}%
%EndExpansion

\section{Discussion}

\label{Dis}

In this work we studied the CKVs of proper (that is the LRS case is
excluded) Bianchi I spacetimes. We have shown that there are only four
families of Bianchi type I spacetimes which admit CKVs. Two of these
families concern conformally flat spacetimes and two non-conformally flat
spacetimes. The non-conformally flat families, to the best of our knowledge,
are new.

One important aspect of these metrics is the symmetry inheritance of the
CKVs by the 4-velocity $u^{a}=\delta _{0}^{a}$ of the comoving observers.
This property is important because it assures that Lie dragging along the
CKVs, fluid flow lines transform onto fluid flow lines thus giving rise to
dynamical conservation laws~\cite%
{maartens-mason-tsamparlis,mason-maartens,coley-tupper1,coley-tupper2,coley-tupper3,coley-tupper4}%
.

The application of the general results of this work to the widely known
Bianchi I\ metrics (\ref{sx3.32}) and (\ref{sx3.35}) has shown that these
spacetimes do not belong to the solutions we have found. More specifically
the Kasner type spacetimes (\ref{sx3.32}) and (\ref{sx3.36}) admit at most a
HV while the Bianchi type I dust solution (\ref{sx3.35}) does not admit even
a HV.

The families of Bianchi I metrics we have found contain many anisotropic
matter solutions which was not possible to be found before due to the
complexity of the conformal equations for Bianchi I spacetimes. It is hoped
that these new solutions will have at least equally interesting properties
as the classical Bianchi I metrics and will make possible the production of
new results mainly at the kinematical level where CKVs play a significant
role.

A final remark concerns the Lie and the Noether point symmetries of
differential equations. Indeed it has been shown that for a general class of
second order partial differential equations the Lie point symmetries are
related to the conformal algebra of the underlying geometry\cite{IJGMMP}.
This class of equations contains among others the heat equation and the
Klein Gordon equation. Therefore one is possible to use the CKVs we have
determined and construct conservation laws or to solve explicitly this type
of differential equations in the corresponding Bianchi I spacetimes.

\begin{acknowledgements} We would like to thank the anonymous referee for helpful comments which have improved the manuscript. AP acknowledge financial support of INFN
\end{acknowledgements}\appendix

\section{Proof of Proposition \protect\ref{prop22}}

\label{appendixA}

In this appendix we give the direct and the inverse proof of Proposition \ref%
{prop22}.

\textbf{Direct Proof: }First recall the decomposition of the curvature
tensor \cite{kramer-stephani}:%
\begin{equation}
R_{abcd}=C_{abcd}+\frac{2}{n-2}\left( g_{c[a}R_{b]d}+g_{d[b}R_{a]c}\right) -%
\frac{R}{(n-1)(n-2)}g_{abcd}  \label{sx2.2}
\end{equation}%
where $g_{abcd}=g_{ac}g_{bd}-g_{ad}g_{bc}$ and the dimension of space is $%
n\geqslant 4.$ Furthermore in a $1+(n-1)$ decomposable space holds that \cite%
{kramer-stephani}:%
\begin{equation}
\overset{n}{R}_{abcd}=\delta _{a}^{\alpha }\delta _{b}^{\beta }\delta
_{c}^{\gamma }\delta _{d}^{\sigma }\overset{n-1}{R}_{\alpha \beta \gamma
\sigma };\qquad \overset{n}{R}_{ab}=\delta _{a}^{\alpha }\delta _{b}^{\beta }%
\overset{n-1}{R}_{\alpha \beta };\qquad \overset{n}{R}=\overset{n-1}{R}.
\label{sx2.4}
\end{equation}%
We consider cases.

Case 1: $n\geqslant 5$

Assume the metric $g_{ab}$ to be conformally flat; then $C_{abcd}=0$.
Replacing $\overset{n}{R}_{abcd},$ $\overset{n}{R}_{ab},$ $\overset{n}{R}$
in (\ref{sx2.2}) and taking into account that $C_{abcd}=0$ we find
\begin{equation}
\overset{n-1}{R}_{\alpha \beta \gamma \delta }=\frac{2}{n-2}\left( g_{\gamma
\lbrack \alpha }\overset{n-1}{R_{\beta ]\delta }}+g_{\delta \lbrack \beta }%
\overset{n-1}{R_{\alpha ]\gamma }}\right) -\frac{\overset{n-1}{R}}{(n-1)(n-2)%
}g_{\alpha \beta \gamma \delta }.  \label{sx2.4.1}
\end{equation}%
where $g_{\alpha \beta \gamma \sigma }$ is defined similarly to $g_{abcd}$.

From (\ref{sx2.2}) we conclude that {\large {\ }}$\overset{n-1}{C}_{\alpha
\beta \gamma \delta }${\large {$=0$}} (because $n-1\geqslant 4)$ therefore
the $n-1$ space is conformally flat. {\large {\ }}Contracting with $%
g^{\alpha \gamma }$ we get:%
\begin{equation}
\overset{n-1}{R}_{\alpha \beta }=\frac{\overset{n-1}{R}}{n-1}g_{\alpha \beta
}  \label{sx2.5}
\end{equation}%
and the $(n-1)$ space is also an Einstein space. We conclude that the $n-1$
space is a space of constant curvature \cite{Eisenhart}.

In order to compute the constant $p$ we insert (\ref{sx2.5}) back to (\ref%
{sx2.2}) and find:%
\begin{equation}
\overset{n-1}{R}_{\alpha \beta \gamma \sigma }=\frac{\overset{n-1}{R}}{%
(n-1)(n-2)}g_{\alpha \beta \gamma \sigma }  \label{sx2.6}
\end{equation}

The $n$ space being conformally flat admits CKVs. According to proposition %
\ref{prop21} these vectors are found from the gradient CKVs of the $(n-1)$
space of the form (\ref{sx2.0b}). Ricci identity for the CKV $\psi _{,\mu }$
gives:%
\begin{equation}
\psi _{|\mu \nu \sigma }-\psi _{|\mu \sigma \nu }=\overset{n-1}{R}_{\sigma
\nu \mu \delta }\psi ^{,\delta }.  \label{sx2.8}
\end{equation}%
Using (\ref{sx2.6}) and (\ref{sx2.0b}) in equation (\ref{sx2.8}) we obtain:%
\begin{equation}
\left[ \frac{\overset{n-1}{R}}{(n-1)(n-2)}+p\right] g_{\alpha \beta \gamma
\delta }\psi ^{,\delta }=0.  \label{sx2.9}
\end{equation}%
from which follows:%
\begin{equation}
\overset{n-1}{R}=-p(n-1)(n-2)  \label{sx2.10}
\end{equation}%
and
\begin{equation*}
p=-\frac{\overset{n-1}{R}}{(n-1)(n-2)}.
\end{equation*}%
Case 2: $n=4$

In this case relation (\ref{sx2.2}) still applies and (\ref{sx2.4.1})
becomes:%
\begin{equation}
\overset{3}{R}_{\alpha \beta \gamma \delta }=\left( g_{\gamma \lbrack \alpha
}\overset{3}{R_{\beta ]\delta }}+g_{\delta \lbrack \beta }\overset{3}{R}%
_{\alpha ]\gamma }\right) -\frac{\overset{3}{R}}{6}g_{\alpha \beta \gamma
\delta }  \label{sx2.10.1}
\end{equation}%
where now the Greek indices take the values 1,2,3. Contracting with $%
g^{\alpha \gamma }$ we find%
\begin{equation}
\overset{3}{R}_{\beta \delta }=\frac{\overset{3}{R}}{3}g_{\beta \delta }
\label{sx2.10.2}
\end{equation}%
which implies that the 3d space is an Einstein space. Although the 3d -
space is an Einstein space of curvature $\overset{3}{R}=const.$ we cannot
conclude that it is a space of constant curvature before we prove that it is
conformally flat. The condition for this is that the Cotton - York tensor
\begin{equation*}
C_{\beta }^{\alpha }=2\varepsilon ^{\alpha \gamma \delta }\left( \overset{3}{%
R}_{\beta \gamma }-\frac{1}{4}g_{\beta \gamma }\overset{3}{R}\right)
_{;\delta }
\end{equation*}%
vanishes \cite{Eisenhart}. Replacing $\overset{3}{R}_{\beta \delta }$ from (%
\ref{sx2.10.2}) we find%
\begin{equation}
C_{\beta }^{\alpha }=\frac{1}{6}\varepsilon ^{\alpha \gamma \delta }g_{\beta
\gamma }\overset{3}{R}_{;\delta }.  \label{sx2.10.3}
\end{equation}

We replace $\overset{3}{R}_{\beta \delta }$ from (\ref{sx2.10.2}) in (\ref%
{sx2.10.1}) and find%
\begin{equation}
\overset{3}{R}_{\alpha \beta \gamma \delta }=\left( g_{\gamma \lbrack \alpha
}\overset{3}{R}_{\beta ]\delta }+g_{\delta \lbrack \beta }\overset{3}{R}%
_{\alpha ]\gamma }\right) -\frac{\overset{3}{R}}{6}g_{\alpha \beta \gamma
\delta }=\frac{\overset{3}{R}}{6}g_{\alpha \beta \gamma \delta }.
\end{equation}%
Ricci identity for the gradient CKV $\psi _{,\mu }$ gives:%
\begin{equation}
\psi _{|\mu \nu \sigma }-\psi _{|\mu \sigma \nu }=\overset{3}{R}_{\sigma \nu
\mu \delta }\psi ^{,\delta }=-\frac{\overset{3}{R}}{6}g_{\mu \nu \sigma
\delta }\psi ^{,\delta }
\end{equation}%
Using (\ref{sx2.6}) and (\ref{sx2.0b}) in equation (\ref{sx2.8}) we obtain:%
\begin{equation}
\left[ \frac{\overset{3}{R}}{6}+p\right] g_{\alpha \beta \gamma \sigma }\psi
^{,\delta }=0
\end{equation}%
from which follows $\overset{3}{R}_{;\delta }=0$ hence $C_{\beta }^{\alpha
}=0,$ which completes the proof.

Case 3: $n=3$

In this case the space $3-1=2$ is conformally flat and admits gradient CKVs
hence the curvature scalar is a constant and the space is a space of
constant curvature.

\textbf{Inverse Proof: \ }\noindent Suppose the $(n-1)$ space of the $1+(n-1)
$ space (\ref{sx2.1}) is a space of constant curvature. Then it is
conformally flat and by (\ref{sx2.2}),(\ref{sx2.4}) and (\ref{sx2.5}) the $%
1+(n-1)$ space is conformally flat. \

This completes the proof of proposition \ref{prop22}

\section{The conformal algebra of RT and ART spacetimes}

\label{appendixB}

The eight proper CKVs of the RT spacetime are%
\begin{equation}
X_{(k)\mu }=a^{2}A_{k,\mu }~~,~X_{(k)z}=-a^{2}A_{k,z}  \label{sx3.6}
\end{equation}%
\begin{equation}
X_{(k+4)\mu }=a^{2}B_{k,\mu }~,~X_{(k+4)z}=-a^{2}B_{k,z}  \label{sx3.8}
\end{equation}%
where $\mu =t,x,y$ and the corresponding conformal factors are
\begin{equation}
\psi _{X_{k}}=A_{k}~,~\psi _{X_{k+4}}=B_{k}
\end{equation}%
where the fields $A_{k},B_{k}$ are given by the expressions%
\begin{equation}
A_{k}=\cos (\frac{\tau }{a})\left\{ \cosh (\frac{y}{a})\left[ \sin (\frac{z}{%
a}),\cos (\frac{z}{a})\right] ,\sinh (\frac{y}{a})\left[ \sin (\frac{z}{a}%
),\cos (\frac{z}{a})\right] \right\}
\end{equation}%
\begin{equation}
B_{k}=\sin (\frac{\tau }{a})\left\{ \cosh (\frac{x}{a})\left[ \sin (\frac{z}{%
a}),\cos (\frac{z}{a})\right] ,\sinh (\frac{x}{a})\left[ \sin (\frac{z}{a}%
),\cos (\frac{z}{a})\right] \right\} .
\end{equation}

The eight proper CKVs of the ART spacetime are
\begin{equation}
Y_{(k)\mu }=-a^{2}\bar{A}_{k,\mu }~~,~Y_{(k)z}=a^{2}\bar{A}_{k,z}
\label{sx3.13}
\end{equation}%
\begin{equation}
Y_{(k+4)\mu }=-a^{2}\bar{B}_{k,\mu }~,~Y_{(k+4)z}=a^{2}\bar{B}_{k,z}
\label{sx3.15}
\end{equation}%
where $\mu =t,x,y$ and the corresponding conformal factors are
\begin{equation}
\psi _{Y_{k}}=A_{k}~,~\psi _{Y_{k+4}}=B_{k}
\end{equation}%
where the fields $\bar{A}_{k},\bar{B}_{k}$ are
\begin{equation}
\bar{A}_{k}=\cosh (\frac{\tau }{a})\left\{ \cos (\frac{y}{a})\left[ \sinh (%
\frac{z}{a}),\cosh (\frac{z}{a})\right] ,\sin (\frac{y}{a})\left[ \sinh (%
\frac{z}{a}),\cosh (\frac{z}{a})\right] \right\}  \label{sx3.17}
\end{equation}%
\begin{equation}
\bar{B}_{k}=\sinh (\frac{\tau }{a})\left\{ \cos (\frac{x}{a})\left[ \sinh (%
\frac{z}{a}),\cosh (\frac{z}{a})\right] ,\sin (\frac{x}{a})\left[ \sinh (%
\frac{z}{a}),\cosh (\frac{z}{a})\right] \right\} .
\end{equation}%
For easy reference in tables \ref{RTCKVS} and \ref{ARTCKVS} we give the
explicit form of the CKVs for the RT spacetime and the ART spacetime
respectively.

Furthermore, the RT spacetime (\ref{sx3.3}) admits a seven dimensional
Killing algebra, the three vector fields are the KVs $\left\{ \partial
_{x},\partial _{y},\partial _{z}\right\} $ and the four extra KVs are%
\begin{equation*}
\mathbf{\xi }_{4,RT}=\sinh (\frac{y}{a})\cosh (\frac{x}{a})\partial _{\tau
}-\cot (\frac{\tau }{a})\sinh (\frac{y}{a})\cosh (\frac{x}{a})\partial
_{x}+\tan (\frac{\tau }{a})\sinh (\frac{y}{a})\cosh (\frac{x}{a})\partial
_{y}
\end{equation*}%
\begin{equation*}
\xi _{5,RT}=\sinh (\frac{y}{a})\sinh (\frac{x}{a})\partial _{\tau }-\cot (%
\frac{\tau }{a})\sinh (\frac{y}{a})\sinh (\frac{x}{a})\partial _{x}+\tan (%
\frac{\tau }{a})\sinh (\frac{y}{a})\sinh (\frac{x}{a})\partial _{y}
\end{equation*}%
\begin{equation*}
\xi _{6,RT}=\cosh (\frac{y}{a})\cosh (\frac{x}{a})\partial _{\tau }-\cot (%
\frac{\tau }{a})\cosh (\frac{y}{a})\cosh (\frac{x}{a})\partial _{x}+\tan (%
\frac{\tau }{a})\cosh (\frac{y}{a})\cosh (\frac{x}{a})\partial _{y}
\end{equation*}%
\begin{equation*}
\xi _{7,RT}=\cosh (\frac{y}{a})\sinh (\frac{x}{a})\partial _{\tau }-\cot (%
\frac{\tau }{a})\cosh (\frac{y}{a})\sinh (\frac{x}{a})\partial _{x}+\tan (%
\frac{\tau }{a})\cosh (\frac{y}{a})\sinh (\frac{x}{a})\partial _{y}
\end{equation*}

Similarly for the ART spacetime (\ref{sx3.4}), the four extra KVs are%
\begin{equation*}
\mathbf{\xi }_{4,ART}=\sin (\frac{y}{a})\cosh (\frac{x}{a})\partial _{\tau
}-\coth (\frac{\tau }{a})\sin (\frac{y}{a})\sinh (\frac{x}{a})\partial
_{x}+\tanh (\frac{\tau }{a})\cos (\frac{y}{a})\cosh (\frac{x}{a})\partial
_{y}
\end{equation*}%
\begin{equation*}
\mathbf{\xi }_{5,ART}=\sin (\frac{y}{a})\sinh (\frac{x}{a})\partial _{\tau
}-\coth (\frac{\tau }{a})\sin (\frac{y}{a})\cosh (\frac{x}{a})\partial
_{x}+\tanh (\frac{\tau }{a})\cos (\frac{y}{a})\sinh (\frac{x}{a})\partial
_{y}
\end{equation*}%
\begin{equation*}
\mathbf{\xi }_{6,ART}=\cos (\frac{y}{a})\cosh (\frac{x}{a})\partial _{\tau
}-\coth (\frac{\tau }{a})\cos (\frac{y}{a})\sinh (\frac{x}{a})\partial
_{x}-\tanh (\frac{\tau }{a})\sin (\frac{y}{a})\cosh (\frac{x}{a})\partial
_{y}
\end{equation*}%
\begin{equation*}
\mathbf{\xi }_{7,ART}=\cos (\frac{y}{a})\sinh (\frac{x}{a})\partial _{\tau
}-\coth (\frac{\tau }{a})\cos (\frac{y}{a})\cosh (\frac{x}{a})\partial
_{x}-\tanh (\frac{\tau }{a})\sin (\frac{y}{a})\sinh (\frac{x}{a})\partial
_{y}
\end{equation*}

%TCIMACRO{\TeXButton{B}{\begin{table}[tbp] \centering}}%
%BeginExpansion
\begin{table}[tbp] \centering%
%EndExpansion
\caption{Proper CKVs of the RT spacetime (\ref{sx3.3})}%
\begin{tabular}{cccccc}
\hline\hline
$\mathbf{X}$ & $\mathbf{X}_{\tau }$ & $\mathbf{X}_{x}$ & $\mathbf{X}_{y}$ & $%
\mathbf{X}_{z}$ & \textbf{Conformal factor }$\mathbf{\psi }$ \\ \hline
$X_{1}$ & $a\sin \left( \frac{\tau }{a}\right) \cosh \left( \frac{y}{a}%
\right) \sin \left( \frac{z}{a}\right) $ & $0$ & $\frac{a\sinh \left( \frac{y%
}{a}\right) \sin \left( \frac{z}{a}\right) }{\cos \left( \frac{\tau }{a}%
\right) }$ & $-a\cos \left( \frac{\tau }{a}\right) \cosh \left( \frac{y}{a}%
\right) \cos \left( \frac{z}{a}\right) $ & $\cos \left( \frac{\tau }{a}%
\right) \cosh \left( \frac{y}{a}\right) \sin \left( \frac{z}{a}\right) $ \\
$X_{2}$ & $a\sin \left( \frac{\tau }{a}\right) \cosh \left( \frac{y}{a}%
\right) \cos \left( \frac{z}{a}\right) $ & $0$ & $\frac{a\sinh \left( \frac{y%
}{a}\right) \cos \left( \frac{z}{a}\right) }{\cos \left( \frac{\tau }{a}%
\right) }$ & $a\cos \left( \frac{\tau }{a}\right) \cosh \left( \frac{y}{a}%
\right) \sin \left( \frac{z}{a}\right) $ & $\cos \left( \frac{\tau }{a}%
\right) \cosh \left( \frac{y}{a}\right) \cos \left( \frac{z}{a}\right) $ \\
$X_{3}$ & $a\sin \left( \frac{\tau }{a}\right) \sinh \left( \frac{y}{a}%
\right) \sin \left( \frac{z}{a}\right) $ & $0$ & $\frac{a\cosh \left( \frac{y%
}{a}\right) \sin \left( \frac{z}{a}\right) }{\cos \left( \frac{\tau }{a}%
\right) }$ & $-a\cos \left( \frac{\tau }{a}\right) \sinh \left( \frac{y}{a}%
\right) \cos \left( \frac{z}{a}\right) $ & $\cos \left( \frac{\tau }{a}%
\right) \sinh \left( \frac{y}{a}\right) \sin \left( \frac{z}{a}\right) $ \\
$X_{4}$ & $a\sin \left( \frac{\tau }{a}\right) \sinh \left( \frac{y}{a}%
\right) \cos \left( \frac{z}{a}\right) $ & $0$ & $\frac{a\cosh \left( \frac{y%
}{a}\right) \cos \left( \frac{z}{a}\right) }{\cos \left( \frac{\tau }{a}%
\right) }$ & $a\cos \left( \frac{\tau }{a}\right) \sinh \left( \frac{y}{a}%
\right) \sin \left( \frac{z}{a}\right) $ & $\cos \left( \frac{\tau }{a}%
\right) \sinh \left( \frac{y}{a}\right) \cos \left( \frac{z}{a}\right) $ \\
$X_{5}$ & $-a\cos \left( \frac{\tau }{a}\right) \cosh \left( \frac{x}{a}%
\right) \sin \left( \frac{z}{a}\right) $ & $\frac{a\sinh \left( \frac{x}{a}%
\right) \sin \left( \frac{z}{a}\right) }{\sin \left( \frac{\tau }{a}\right) }
$ & $0$ & $-a\sin \left( \frac{\tau }{a}\right) \cosh \left( \frac{x}{a}%
\right) \cos \left( \frac{z}{a}\right) $ & $\sin \left( \frac{\tau }{a}%
\right) \cosh \left( \frac{x}{a}\right) \sin \left( \frac{z}{a}\right) $ \\
$X_{6}$ & $-a\cos \left( \frac{\tau }{a}\right) \cosh \left( \frac{x}{a}%
\right) \cos \left( \frac{z}{a}\right) $ & $\frac{a\sinh \left( \frac{x}{a}%
\right) \cos \left( \frac{z}{a}\right) }{\sin \left( \frac{\tau }{a}\right) }
$ & $0$ & $a\sin \left( \frac{\tau }{a}\right) \cosh \left( \frac{x}{a}%
\right) \sin \left( \frac{z}{a}\right) $ & $\sin \left( \frac{\tau }{a}%
\right) \cosh \left( \frac{x}{a}\right) \cos \left( \frac{z}{a}\right) $ \\
$X_{7}$ & $-a\cos \left( \frac{\tau }{a}\right) \sinh \left( \frac{x}{a}%
\right) \sin \left( \frac{z}{a}\right) $ & $\frac{a\cosh \left( \frac{x}{a}%
\right) \sin \left( \frac{z}{a}\right) }{\sin \left( \frac{\tau }{a}\right) }
$ & $0$ & $-a\sin \left( \frac{\tau }{a}\right) \sinh \left( \frac{x}{a}%
\right) \cos \left( \frac{z}{a}\right) $ & $\sin \left( \frac{\tau }{a}%
\right) \sinh \left( \frac{x}{a}\right) \sin \left( \frac{z}{a}\right) $ \\
$X_{8}$ & $-a\cos \left( \frac{\tau }{a}\right) \sinh \left( \frac{x}{a}%
\right) \cos \left( \frac{z}{a}\right) $ & $\frac{a\cosh \left( \frac{x}{a}%
\right) \cos \left( \frac{z}{a}\right) }{\sin \left( \frac{\tau }{a}\right) }
$ & $0$ & $a\sin \left( \frac{\tau }{a}\right) \sinh \left( \frac{x}{a}%
\right) \sin \left( \frac{z}{a}\right) $ & $\sin \left( \frac{\tau }{a}%
\right) \sinh \left( \frac{x}{a}\right) \cos \left( \frac{z}{a}\right) $ \\
\hline\hline
\end{tabular}%
\label{RTCKVS}%
%TCIMACRO{\TeXButton{E}{\end{table}}}%
%BeginExpansion
\end{table}%
%EndExpansion

%TCIMACRO{\TeXButton{B}{\begin{table}[tbp] \centering}}%
%BeginExpansion
\begin{table}[tbp] \centering%
%EndExpansion
\caption{The proper CKVs of the ART spacetime (\ref{sx3.4})}%
\begin{tabular}{cccccc}
\hline\hline
$\mathbf{X}$ & $\mathbf{X}_{\tau}$ & $\mathbf{X}_{x}$ & $\mathbf{X}_{y}$ & $%
\mathbf{X}_{z}$ & \textbf{Conformal factor }$\mathbf{\psi }$ \\ \hline
$X_{1}$ & $a\sinh \left( \frac{\tau }{a}\right) \cos \left( \frac{y}{a}%
\right) \sinh \left( \frac{z}{a}\right) $ & $0$ & $\frac{a\sin \left( \frac{y%
}{a}\right) \sinh \left( \frac{z}{a}\right) }{\cosh \left( \frac{\tau}{%
\alpha }\right) }$ & $a\cosh \left( \frac{\tau }{a}\right) \cos \left( \frac{%
y}{a}\right) \cosh \left( \frac{z}{a}\right) $ & $\cosh \left( \frac{\tau }{a%
}\right) \cos \left( \frac{y}{a}\right) \sinh \left( \frac{z}{a}\right) $ \\
$X_{2}$ & $a\sinh \left( \frac{\tau }{a}\right) \cos \left( \frac{y}{a}%
\right) \cosh \left( \frac{z}{a}\right) $ & $0$ & $\frac{a\sin \left( \frac{y%
}{a}\right) \cosh \left( \frac{z}{a}\right) }{\cosh \left( \frac{\tau}{%
\alpha }\right) }$ & $a\cosh \left( \frac{\tau }{a}\right) \cos \left( \frac{%
y}{a}\right) \sinh \left( \frac{z}{a}\right) $ & $\cosh \left( \frac{\tau }{a%
}\right) \cos \left( \frac{y}{a}\right) \cosh \left( \frac{z}{a}\right) $ \\
$X_{3}$ & $a\sinh \left( \frac{\tau }{a}\right) \sin \left( \frac{y}{a}%
\right) \sinh \left( \frac{z}{a}\right) $ & $0$ & $-\frac{a\cos \left( \frac{%
y}{a}\right) \sinh \left( \frac{z}{a}\right) }{\cosh \left( \frac{\tau}{%
\alpha }\right) }$ & $a\cosh \left( \frac{\tau }{a}\right) \sin \left( \frac{%
y}{a}\right) \cosh \left( \frac{z}{a}\right) $ & $\cosh \left( \frac{\tau }{a%
}\right) \sin \left( \frac{y}{a}\right) \sinh \left( \frac{z}{a}\right) $ \\
$X_{4}$ & $a\sinh \left( \frac{\tau }{a}\right) \sin \left( \frac{y}{a}%
\right) \cosh \left( \frac{z}{a}\right) $ & $0$ & $-\frac{a\cos \left( \frac{%
y}{a}\right) \cosh \left( \frac{z}{a}\right) }{\cosh \left( \frac{\tau}{%
\alpha }\right) }$ & $a\cosh \left( \frac{\tau }{a}\right) \sin \left( \frac{%
y}{a}\right) \sinh \left( \frac{z}{a}\right) $ & $\cosh \left( \frac{\tau }{a%
}\right) \sin \left( \frac{y}{a}\right) \cosh \left( \frac{z}{a}\right) $ \\
$X_{5}$ & $a\cosh \left( \frac{\tau }{a}\right) \cos \left( \frac{x}{a}%
\right) \sinh \left( \frac{z}{a}\right) $ & $\frac{a\sin \left( \frac{x}{a}%
\right) \sinh \left( \frac{z}{a}\right) }{\sinh \left( \frac{\tau }{a}%
\right) }$ & $0$ & $a\sinh \left( \frac{\tau }{a}\right) \cos \left( \frac{x%
}{a}\right) \cosh \left( \frac{z}{\alpha }\right) $ & $\sinh \left( \frac{%
\tau }{a}\right) \cos \left( \frac{x}{a}\right) \sinh \left( \frac{z}{a}%
\right) $ \\
$X_{6}$ & $a\cosh \left( \frac{\tau }{a}\right) \cos \left( \frac{x}{a}%
\right) \cosh \left( \frac{z}{a}\right) $ & $\frac{a\sin \left( \frac{x}{a}%
\right) \cosh \left( \frac{z}{a}\right) }{\sinh \left( \frac{\tau }{a}%
\right) }$ & $0$ & $a\sinh \left( \frac{\tau }{a}\right) \cos \left( \frac{x%
}{a}\right) \sinh \left( \frac{z}{\alpha }\right) $ & $\sinh \left( \frac{%
\tau }{a}\right) \cos \left( \frac{x}{a}\right) \cosh \left( \frac{z}{a}%
\right) $ \\
$X_{7}$ & $a\cosh \left( \frac{\tau }{a}\right) \sin \left( \frac{x}{a}%
\right) \sinh \left( \frac{z}{a}\right) $ & $-\frac{a\cos \left( \frac{x}{a}%
\right) \sinh \left( \frac{z}{a}\right) }{\sinh \left( \frac{\tau }{a}%
\right) }$ & $0$ & $a\sinh \left( \frac{\tau }{a}\right) \sin \left( \frac{x%
}{a}\right) \cosh \left( \frac{z}{\alpha }\right) $ & $\sinh \left( \frac{%
\tau }{a}\right) \sin \left( \frac{x}{a}\right) \sinh \left( \frac{z}{a}%
\right) $ \\
$X_{8}$ & $a\cosh \left( \frac{\tau }{a}\right) \sin \left( \frac{x}{a}%
\right) \cosh \left( \frac{z}{a}\right) $ & $-\frac{a\cos \left( \frac{x}{a}%
\right) \cosh \left( \frac{z}{a}\right) }{\sinh \left( \frac{\tau }{a}%
\right) }$ & $0$ & $a\sinh \left( \frac{\tau }{a}\right) \sin \left( \frac{x%
}{a}\right) \sinh \left( \frac{z}{\alpha }\right) $ & $\sinh \left( \frac{%
\tau }{a}\right) \sin \left( \frac{x}{a}\right) \cosh \left( \frac{z}{a}%
\right) $ \\ \hline\hline
\end{tabular}%
\label{ARTCKVS}%
%TCIMACRO{\TeXButton{E}{\end{table}}}%
%BeginExpansion
\end{table}%
%EndExpansion

\end{document}